\documentclass[12pt]{article}
\usepackage{graphicx}
\usepackage{psfrag}
\usepackage{epstopdf}

\begin{document}

\begin{titlepage}
\null\vspace{-62pt}

\pagestyle{empty}
\begin{center}

\vspace{1.0truein} {\Large\bf Radiative generation of metastable minima in a scalar-fermion model}

\vspace{1in}
{\large\bf Dimitrios Metaxas} \\
\vskip .4in
{\it Department of Physics,\\
National Technical University of Athens,\\
Zografou Campus, 15780 Athens, Greece\\
metaxas@central.ntua.gr}\\

\vspace{0.5in}

\vspace{.5in}
\centerline{\bf Abstract}

\baselineskip 18pt
\end{center}

I consider a theory of a real scalar and a fermion field, with a Yukawa interaction and a potential term that admits two degenerate minima at the tree level. The quantum vacuum energy difference between these two vacua can be calculated using the renormalization group improved effective potential, and gives a finite, nonzero result, dependent on the relative strength of the scalar and the Yukawa interactions.

\end{titlepage}

\newpage
\pagestyle{plain}
\setcounter{page}{1}
\newpage

\section{Introduction}

The effective action and especially the effective potential term have proven to be very useful tools in order to investigate the vacuum structure
of quantum field theories \cite{cw}. They become particularly important in questions of stability of the Standard Model vacuum,
in phase transitions in cosmology or high temperature field theory, and in the associated phenomena of tunneling or ``slow-roll'' evolution of
the unstable or metastable vacua.

 An intriguing question is that of the absolute value of the vacuum energy, or ``cosmological constant'' term associated with these processes. In flat space-time, the value of the absolute minimum of the effective potential is not a measurable quantity,
it should be included, however, for a proper renormalization group treatment. Also,
potential energy differences between different vacua (local minima of the effective potential) are, in principle, measurable.

Here, I give an example of a potential term that has two degenerate minima at tree level, because of quantum effects, however,
after renormalization and resummation of infrared divergencies, they become inequivalent, with a finite, nonzero potential energy difference. A fermion field is included that is originally massless in one minimum, so that renormalization conditions are imposed
at a nonzero value of the scalar field. As a result, the absolute values of the two minima are shifted, depending on the relative values of the scalar and Yukawa couplings. 
The vacuum energy difference between the two minima
 is a quantum radiative effect and can be estimated with the help of the renormalization group
improved effective potential.

In Sec.~2, I give a review of the question of vacuum energy in quantum field theory in flat space-time and
its relation to symmetry considerations. In Sec.~3, I discuss the model which is asymmetric because of quantum effects,
and in Sec.~4, I conclude with some more comments.

\section{Effective potential and vacuum energy}

I will start by reviewing the problem of the vacuum energy for
renormalizable quantum field theories, in four-dimensional flat spacetime, that contain a scalar field, $\phi$, which is endowed, at tree level, with a standard kinetic term and a general potential term, $U(\phi)$, which is bounded below.

If the potential term  at hand has a single minimum (vacuum) at $\phi=\phi_{\rm min}$ then quantization can be performed around it after expanding $U(\phi)=U(\phi_{\rm min}) + \frac{1}{2} U''(\phi_{\rm min}) (\phi-\phi_{\rm min})^2 + \cdots$, discarding the constant term, using the quadratic term to describe a scalar excitation of mass $m$ around the minimum, with $m^2 =U''(\phi_{\rm min})$, and treating the higher order terms in perturbation theory as interactions with the respective coupling constants.

The constant term, also called the vacuum energy term, along with the mass and the coefficients of the higher order interactions have no meaning at this point; they are called bare terms and get regularized by (infinite) multiplications or subtractions, along with a similar treatment of the kinetic term in the usual process of renormalization.

Associated with this procedure are two parameters, both with dimensions of mass: $\Lambda$,
which is used in order to cutoff divergent expressions,
and $\mu$, that sets the scale in which the physical parameters of the theory, masses and coupling constants
are defined or measured.
Then one proceeds by calculating, order by order in the perturbation expansion, the various Green's functions of the theory as well as the related functional expressions of the effective action with the corresponding effective potential \cite{cw}.

The cutoff, $\Lambda$, was just a mathematical convention and should be absent from any final result of these calculations.
The theory is defined by specifying the values of the masses and the coupling constants at a scale $\mu$;
although the Green's functions and the effective action depend on the scale,
any physical result derived from them should be $\mu$-independent. For example: one may measure and define the masses and coupling constants of the theory using certain scattering experiments at a ``reference'' scale $\mu_{\rm ref} = 1$GeV. Then one may predict and measure the outcomes of any experiment at any other scale, say, $\mu_{\rm exp}=10$GeV. The result should be the same with what one would have obtained after having used a different $\mu_{\rm ref}$ to start with. This is embodied in the renormalization group formalism and is expressed mathematically by the fact that the total derivative of any physical quantity with respect to $\mu$, given by the sum of the various partial derivatives, must vanish.

We see immediately the reason why the constant, vacuum energy, term was discarded: there is no physical process or experiment that depends on it; it can be set to zero, or any other value if one is not worried about the semiclassical expansion around an infinite constant. Once this is done (here I will consider it set to zero) there is no
prediction for a different value, nor can there be any process to verify such a prediction.
If one wants to use the renormalization group formalism consistently, however, one must take care of the constant term too, that is, in our case, subtract its value at the minimum at any level in the perturbation expansion of the effective potential \cite{rg1}.

When the theory under consideration is coupled to gravity, whether the latter is considered at the classical level or quantized, the value of the vacuum energy becomes a physical observable that can be measured in the cosmological expansion rate and contributes to the cosmological constant \cite{cc1}. The quantum theory of gravity is not renormalizable; it can be viewed as an effective quantum field theory \cite{eff}, with a limited range of predictability, as all effective quantum field theories, and its implications will not be considered here. As far as renormalizable quantum field theories in flat space-time are concerned, there can be no prediction for the vacuum energy defined as the value of the renormalized effective potential at its minimum.

It is sometimes argued that the sum of the zero-point energies of the field modes at the minimum contributes a factor of
\begin{equation}
\frac{1}{4\pi^2}\int_0^{\Lambda}
dk\,k^2\,\sqrt{k^2+m^2}
\approx \frac{\Lambda^4}{16\pi^2}
\label{arg1}
\end{equation}
when a momentum cutoff regularization scheme is employed,
or
\begin{equation}
\frac{\mu^{4-d}}{(2\pi)^{(d-1)}}
\frac{1}{2}
\int
d^{d-1}k\sqrt{k^2+m^2}
\approx
\frac{m^4}{64 \pi^2} \ln \left(\frac{m^2}{\mu^2}\right)
\label{arg2}
\end{equation}
when dimensional regularization and minimal subtraction prescription are performed.
In (\ref{arg2}), a fermion field would have given a contribution with the opposite sign, involving, of course, the fermion
mass at the minimum.
The cut-off, $\Lambda$, is usually considered to be related to the Planck or a Grand Unified Theory (GUT) scale,
and  the scale $\mu$ to the radiation associated with the supernova observations or the Cosmic Microwave Background \cite{cc2}.

Although these expressions are suggestive of contributions to the vacuum energy
that drive it away from a zero value
when nonrenormalizable interactions such as gravity are considered,
they can hardly be considered as a prediction of a renormalizable quantum field theory.
Higher energy scales, such as the GUT scale, may or may not leave an imprint on processes at the electroweak scale
depending on the details of the decoupling procedure, none of the contributions, however, may depend explicitly on the cutoff in a way implied by (\ref{arg1}). As far as the expression in (\ref{arg2}) is concerned, one also sees that it cannot, by itself, correspond to a well-defined prediction; it is rather a one-loop result that should be subtracted if the perturbation expansion around the vacuum is to be done consistently.

Let us now consider the case where the potential energy term, $U(\phi)$, has, besides the global minimum at $\phi_{\rm min}$, a second, local minimum at $\phi_{\rm met}$, such that $U(\phi_{\rm met}) > U(\phi_{\rm min})$. This local minimum corresponds to a ``false'', metastable vacuum, and the energy difference between the two vacua is a physical observable that can, in principle, be measured if an appropriate metastable state is prepared.
The perturbation expansion of the effective potential must account for this fact; the renormalization group equation \cite{rg1}  will ensure that the vacuum energy difference can be consistently defined, the value of the ``true'' vacuum energy, however, is still undetermined and can be set to zero. Only the energy difference between the two vacua is a meaningful, physical quantity.

This vacuum energy difference is also an input of the theory, much like the various masses and coupling constants; it is not a prediction of the quantum field theory.
Similar considerations apply when the global minimum of the potential was not present at tree level but was induced by radiative, quantum effects \cite{cw}. The dimensionful parameter that defines the location of the absolute minimum and its related energy difference with respect to the metastable one is again an input of the theory, although ``camouflaged'' at the tree level.
As an additional, important note for these cases, one should mention that the energy of a metastable state has an imaginary part that is related to the rate of its decay \cite{tunnel}; this is a nonperturbative effect, however, and will not show up at any level of the perturbation expansion.

One may also consider a theory where the potential energy term at tree level has a discrete or continuous family of degenerate minima that are related by a symmetry. Two simple examples that one can have in mind involve a complex scalar field with a ``Mexican-hat'' potential, or a real scalar field with a ``reflection'' symmetry of
$\phi\rightarrow -\phi$.
Quantization can again be performed by picking one of the minima, thus breaking the symmetry, and following the same procedure as above. The value of the potential at the minimum is again undefined and can be consistently set to zero. Once this is done, by symmetry considerations, the value of the renormalized potential at any other minimum will be zero as well.

Finally, coming to the case that is relevant to the present work, one can imagine the case of a potential term with a set of degenerate minima that have the same value of the energy at tree level but are not otherwise related by any symmetry. A simple example would be a potential with two minima at $\phi_1$ and $\phi_2$, such that $U(\phi_1)=U(\phi_2)$ but $U''(\phi_1)\neq U''(\phi_2)$. Then the elementary excitations around each minimum would have different masses. If one were to pick one minimum, say $\phi_1$, to quantize the theory, all the subtractions described before would have to be performed at this point, and the difference of terms such as (\ref{arg2}) around the two minima should give a finite, possibly nonzero result for $\phi_2$. This would be a definite prediction for the energy of the second vacuum, similar to well-known phenomena like the Casimir effect \cite{casimir}.
Obviously, it is not possible to have a renormalizable quantum field theory in four dimensions with such a potential term at tree level (it is interesting, nevertheless , that the effective potential in the Standard Model allows for the  possibility of a second minimum, other than the one in the electroweak scale, close to the Planck scale and degenerate in energy \cite{higgs}). However, there are other cases where asymmetries between classically degenerate vacua may develop and this is investigated further
in the next Section.

\section{An asymmetric theory}

In order to examine a simple case of a generated asymmetry, I will consider here a theory with a real scalar and a fermion field with a Yukawa interaction and the Lagrangian:
\begin{equation}
{\cal L}=\frac{1}{2}(\partial\phi)^2 - U_0(\phi)
         +i \bar{\psi} \partial\!\!\!\slash \psi -g_0\phi\bar{\psi}\psi,
\label{lll}
\end{equation}
where the potential term,
\begin{equation}
U_0(\phi)=\frac{\lambda_0}{4!}\phi^2(\phi-\phi_0)^2,
\label{pot1}
\end{equation}
has two degenerate minima at $\phi=0$ and $\phi=\phi_0$.
The fermion acquires a mass, $m_{\rm f}=g\phi_0$, around the second minimum, while it is massless around the first. The masses of the scalar excitations may be the same around the two vacua,
\begin{equation}
U_0''(0)=U_0''(\phi_0)=\frac{\lambda_0}{12}\phi_0^2,
\end{equation}
since renormalization involves a scale, $\mu$, however,
and because of the different fermion masses in the two minima,  there is a resulting asymmetry between the zero and the nonzero vacuum.

The effective potential at one loop, after dimensional regularization, is
given by the well-known expression
\begin{equation}
U_{\rm eff}(\phi)=U_0(\phi)+
\frac{1}{64\pi^2}\left[ (U_0'')^2\left(\ln\frac{U_0''}{\mu^2}-\frac{3}{2}\right)
-4g_0^4\phi^4 \left( \ln \frac{g_0^2\phi^2}{\mu^2}-\frac{3}{2}\right)\right].
\label{eff1}
\end{equation}
One can see that all counterterms, from linear to quartic, are generated,
and renormalization conditions for the quartic scalar interaction cannot be taken at $\phi=0$ because of the infrared divergencies.
As far as the fermion field is concerned, we impose the renormalization condition
that the fermion mass is zero at the origin, $m_f(\phi=0)=0$,  
as in the original Lagrangian, (\ref{lll}).
Then, when renormalization conditions for the scalar field are imposed at a nonzero value of the scalar field for the one-loop expression,
one will get a nonzero value for the vacuum energy difference between the two resulting vacua.

However, the one-loop result is not reliable because
 the renormalization scale, $\mu$, cannot be chosen so that both the scalar and the
fermion loops are suppressed.
In order to examine the quantum effective potential one can use the renormalization group improvement,
that is also important in questions of stability \cite{rg1,rg2}.

In the renormalization group treatment, with a running scale
\begin{equation}
\mu^2(t)=\mu_0^2 \, e^{2 t},
\end{equation}
we can consider the general effective potential
\begin{equation}
U(\phi, t)=D(t) \phi +\frac{1}{2} m^2(t) \phi^2 +\frac{1}{6} A(t) \phi^3 + 
              \frac{1}{4!}\lambda(t) \phi^4,
\end{equation}
with the boundary conditions
\begin{equation}
D(0)=0,\,\,m^2(0)=\frac{\lambda_0}{12}\phi_0^2,\,\,
A(0)=-\frac{\lambda_0\phi_0}{2},\,\,\lambda(0)=\lambda_0,\,\,
g^2(0)=g_0^2
\end{equation}
and the beta- and gamma- functions
\begin{equation}
\beta_\lambda=c\, (3\lambda^2+8 g^2 \lambda - 48 g^4)
\end{equation}
\begin{equation}
\beta_{g^2}=c \, 10 g^4
\end{equation}
\begin{equation}
\beta_{m^2}=c\,(\lambda m^2 + A^2 + 4 m^2 g^2)
\end{equation}
\begin{equation}
\beta_A = c \, A(3\lambda +6 g^2)
\end{equation}
\begin{equation}
\beta_D= c\,(A m^2 + 2 g^2 D)
\end{equation}
\begin{equation}
\gamma_\phi =c \, 2 g^2
\end{equation}
with $c=1/16\pi^2$.
Then, the general expression for the renormalization group improved effective potential is
\begin{eqnarray}
U_{\rm RGI}(\phi)=\xi^4(t)\Biggl[ \Omega(t) &+& U(\phi,t) +
                                                    \frac{c}{4}\Biggl( U''(\phi,t)
                                                   \left(\ln\frac{U''}{\mu^2(t)}-\frac{3}{2}\right)
\label{vrgi1}
\\ \nonumber
&-&4 g^4(t) \phi^4 \left( \ln\frac{g^2(t)\phi^2}{\mu^2(t)} -\frac{3}{2} \right) \Biggr) \Biggr],
\end{eqnarray}
with 
\begin{equation}
\xi (t) = \exp \left(- \int_0^t \gamma_\phi(t') dt' \right)
\end{equation}
and the evolution equations
\begin{equation}
\frac{d \lambda_i}{dt} = \bar{\beta}_{\lambda_i},
\end{equation}
where $\lambda_i = \lambda, A, m^2, D, g^2$ are the various couplings and
\begin{equation}
\bar{\beta}_{\lambda_i} = \beta_{\lambda_i} + \delta_i \lambda_i \gamma_\phi,
\end{equation}
where $\delta_i$ is the mass dimension of the respective coupling (no summation over $i$ is implied in the last equation).
$\Omega(t)$ in (\ref{vrgi1}) is the running ``cosmological constant'' or vacuum energy term \cite{rg1, rg2},
that satisfies an analogous equation
\begin{equation}
\frac{d\Omega}{dt}=\frac{c}{2} m^4,
\label{ccrg1}
\end{equation}
with  boundary condition $\Omega(0)=0$.

Now, one may choose the value of the scale $\mu(t)$ so as to diminish the contributions of higher loops
for certain terms; in order to have a more explicit expression for $t$, I will take
\begin{equation}
\mu^2 (t) =\mu_0^2 \, e^{2 t} = g_0^2 \, \phi^2 e^{-\frac{3}{2}}
\end{equation}
which takes care of the fermion contribution, except for a small remaining term,
since the evolution of $g^2$ is given by
\begin{equation}
g^2(t)=\frac{g_0^2}{1-10 \, c \, g_0^2 \, t}.
\label{rgg2}
\end{equation}
The evolution equation for $\lambda$ can also be solved exactly \cite{rg2} and used to examine the stability
issues at large values of $\phi$. The exact form of the solution depends on the initial values of $\lambda$ and $g^2$
and will not be shown here; the remaining equations can be solved numerically for $t \geq 0$ given the initial conditions.

 For small values of $\phi$, near $0$, one still has problematic contributions
from the scalar loops. However, the behavior of the  system at the infrared is 
different: the massive scalar decouples from the massless fermion, with  the result that,
for $t \leq  0$, the evolution of the dimensionless coupling constant, $\lambda$, is given only by the fermion loop contribution,
\begin{equation}
\lambda(t)=\lambda(0) - 48  c   g^4  t,
\end{equation}
while the remaining dimensionful parameters remain ``frozen'' at their $t=0$ values,
and there is no scalar loop contribution in (\ref{vrgi1}) \cite{rg2, rg3}.

The matching between these two regions will be done at $t=0$,
at a value for the scale, $\mu_0$, and the corresponding
field, $\tilde{\phi}$, such that
$\mu_0^2 = U''(\tilde{\phi})$.

For the quantity $\Omega(t)$, I will use the approximate solution to (\ref{ccrg1}),
\begin{equation}
\Omega(t)=-\frac{c}{4}m^4(t)\left( \ln\frac{m^2(t)}{\mu^2(t)}-\frac{3}{2}\right),
\end{equation}
with the value at $t=0$ subtracted, so that $\Omega=0$ for $t \leq 0$.
It is possible to look for an exact solution of (\ref{ccrg1}) that is also 
explicitly $\mu$-independent, the analysis, however, becomes more complicated
and, in our problem, the contribution of this $\Omega$-term anyway
turns out to be subleading, except for cases where
the one-loop expression also needs improvement.

After solving the evolution equations with the corresponding boundary conditions,
and doing the matching at $t=0$ as described before, I show the results
for the renormalization group improved effective potential, for various initial values,
$\lambda_0$ and $g^2_0$ of the couplings in the following Figures
(all dimensionful quantities are in units of appropriate powers of $\phi_0$).

In Fig.~1, I show the renormalization group improved effective potential, $U_{\rm RGI}(\phi)$, as a function of $\phi$ 
for $g_0^2=0.1$ and three different values for $\lambda_0 =0.1, 0.4, 0.8 $ from 
top to bottom. The matching at $t=0$ is done at the value of $\tilde{\phi}=0.145, 0.183, 0.195$
respectively (in units of $\phi_0$), and the second minimum is at $\phi=1.143, 1.028, 1.010$.
The potential energy difference between the two vacua is $0.0003, 0.0002, 0.0001$ respectively
(in units of $\phi_0^4$)
and the value of the $\Omega$ term at the second minimum is at least one order of magnitude smaller.

In Fig.~2, I show the renormalization group improved effective potential, $U_{\rm RGI}(\phi)$, as a function of $\phi$ 
for $\lambda_0 =0.8$ and two different values for $g_0^2=0.05, 0.01$ from top to bottom.
The matching is done for $\tilde{\phi}=0.202, 0.209$ respectively, and the second minimum is almost exactly at $\phi=1$.
The potential energy difference between the two vacua at $\phi=0, \phi_0$ is of the order of $10^{-5}$, 
and the value of $\Omega$ at $\phi_0$ is of the same order of magnitude, both much smaller than the other characteristic
scales of $U(\phi)$. It is possible to look 
for an exact expression for $\Omega$; however, for these values of the couplings (a Yukawa coupling much smaller than 
the scalar coupling) one can see that the one-loop renormalization group treatment receives additional, comparable corrections from
higher loop contributions and the analysis needs to be modified \cite{rg2, rg3}.
Another way to see this is that, for much smaller Yukawa couplings, the fermion at the nonzero minimum has a much smaller mass
than the scalar which effectively decouples as in the case with the massless fermion at $\phi=0$.
In conclusion, the results of Fig.~2 are not expected to be numerically exact, they were included here, however, in order to show
that they agree with the general tendency that the vacuum energy difference is diminished as the relative strength
of the scalar to Yukawa coupling grows. It is also interesting that the $\Omega$ term is of a smaller order of magnitude in the results of Fig.~1,
and the vacuum energy difference there is essentially due to the renormalization group running of the other parameters, whereas
it becomes important for values of the couplings as in Fig.~2, where a more complete treatment of the renormalization group
with different scales is needed.

I should mention at this point that I show, in both Figures, the real part of the effective potential.
 As is well known, there is a region in field space where the final expression for the one-loop effective potential has an imaginary part \cite{im}. It is the region where
$U''(\phi)<0$, and one has to be more careful when deriving physical results associated with this part of the field space. Our areas of interest, however, near $\phi=0$ and $\phi=\phi_0$, have no overlap with the problematic region in this case; also the matching point, $\tilde{\phi}$, is outside this region, obviously, by its definition from
$\mu_0^2=U''(\tilde{\phi})$.

The results of the analysis presented here are consistent with the treatment of the scalar-fermion system in
\cite{rg2} that was focused more on the question of stability.
There, it was shown that the evolution of $\lambda/g^2$ 
 has an infrared and an ultraviolet
fixed point, $w_+=4.35, w_-=-3.68$ respectively, and for 
values of the couplings such that $\lambda_0/g_0^2 < w_+$,  the couplings will flow to the ultraviolet fixed point,
leading to an instability at large values of the field, an effective potential unbounded from below.
This is the case for the first two graphs in Fig.~1, although the corresponding large values of the field have not been shown here.
It is interesting, however, that both effects of a vacuum energy difference and a vacuum instability can be seen in a simple model,
and that they are, in fact related, since they both depend on the relative strength of the scalar and Yukawa interactions.

\section{Comments}

Now we can proceed to discuss the implications of the results of the previous Section.
The fact that the two, classically degenerate, vacua are energetically inequivalent because of quantum corrections, gives this simple model a structure that is richer than expected. The vacuum with higher energy, $\phi=0$ in this case, becomes metastable, although it was classically stable. One can accordingly calculate its rate of decay; the appropriate formalism is related to the results of \cite{ejw2} although the physical situation here is different. Since the vacuum energy difference is a quantum effect, the result for this vacuum decay rate is extremely small, being proportional to the exponential of minus the ``bounce'' action.
In fact, since the system, for certain values of the couplings, may also develop an instability at much larger values of the field, there are additional modes of vacuum decay 
that can be investigated in the same problem, with a similar manner.
 It would be interesting, therefore, as a problem for further research, to study the evolution of the vacua and the effective potential in a finite temperature and cosmological setting in this or related problems where the breaking or the lack of symmetry play an important role \cite{jean}.

One may also consider the possibility of a ``landscape'' of vacua, a large number of which are degenerate, with zero energy at the classical level or even after some quantum corrections have been taken into account. Unless they are all related by the same symmetries, it does not seem possible to have zero energy in all of them when higher order quantum effects are considered, and the energy difference between two adjacent vacua, if one literally translates the results obtained here, would be related to the relative strength of the various coupling constants times their distance in field space.
It is an attractive scenario which states that if the value of the vacuum energy of a particular minimum is fixed by some reason to be zero, the value of the vacuum energy for any nearby minimum will be a suppressed and calculable number.

One frequently encounters the problem, however, that some of the interactions that are involved, in this or other
physically important
situations, are nonrenormalizable, the most important example being the gravitational interaction; when these are regarded as effective quantum field theories \cite{eff}, instead of a coupling constant expansion that was the basic tool of renormalizable theories, one now has an expansion in powers of the energy, and it is possible that well-defined results
for the vacuum energy or energy difference exist in these situations as well.
It would be interesting, therefore, to consider the results of similar problems in effective quantum field theories .

\vspace{0.4in}

\centerline{\large\bf  Acknowledgements}
\noindent
This work was completed at the National Technical University of Athens. I would like to thank the people of the Physics Department for their support.

\begin{figure}
\centering
\includegraphics[width=90mm]{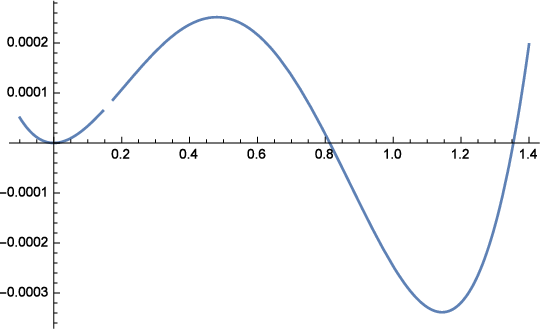}
\includegraphics[width=90mm]{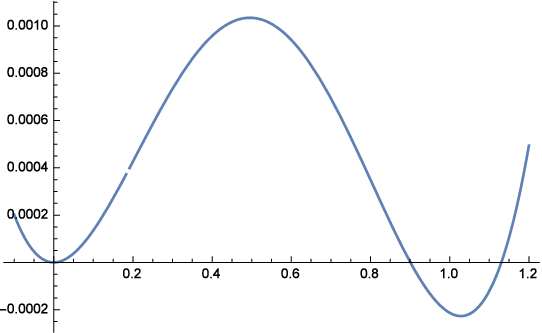}
\includegraphics[width=90mm]{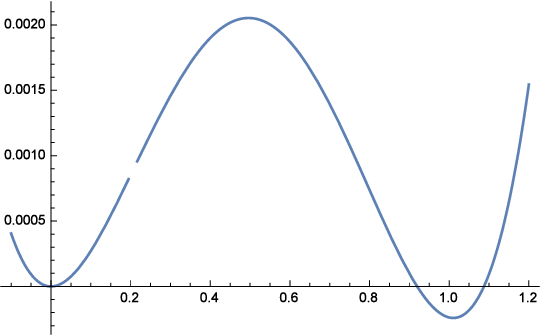}
\caption{I show the one-loop, renormalization group improved effective potential, $U_{\rm RGI}(\phi)$, as a function of $\phi$ 
for $g_0^2=0.1$ and three different values for $\lambda_0 =0.1, 0.4, 0.8 $ from 
top to bottom. The horizontal axis is in units of $\phi_0$ and the vertical axis is in units of $\phi_0^4$.
In the first two cases, the potential becomes unstable (unbounded below) at much larger values of $\phi$, not shown here.}
\end{figure}

\begin{figure}
\centering
\includegraphics[width=90mm]{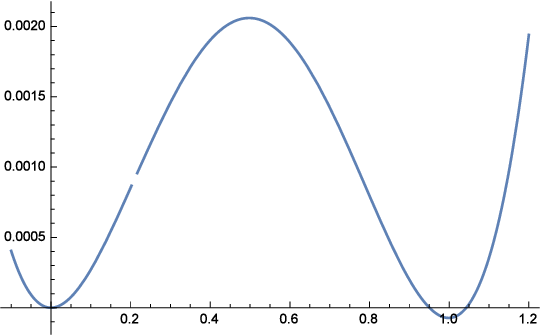}
\includegraphics[width=90mm]{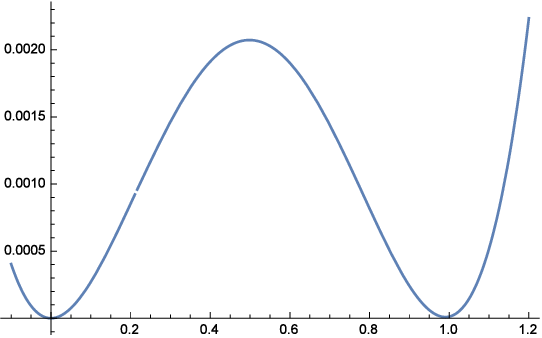}
\caption{I show the one-loop, renormalization group improved effective potential, $U_{\rm RGI}(\phi)$, as a function of $\phi$ 
for $\lambda_0 =0.8$ and two different values for $g_0^2=0.05, 0.01$ from 
top to bottom. The horizontal axis is in units of $\phi_0$ and the vertical axis is in units of $\phi_0^4$.
}
\end{figure}

\end{document}